\documentstyle[prl,tighten,aps,epsf,graphics,preprint,amstex]{revtex}
\newcommand{\ket}[1]{\left | \, #1 \right \rangle}
\newcommand{\bra}[1]{\left \langle \, #1 \right | }
\newcommand{\id}{{\sf 1 \hspace{-0.3ex} \rule{0.1ex}{1.52ex}
\rule[-.01ex]{0.3ex}{0.1ex}}}
\newcommand{\proy}{I\!\!P}

\begin{document}
\draft

\title{Measuring the entanglement of bipartite pure states}
\author{J. M. G. Sancho$^1$ and S. F. Huelga$^{1,2}$}
\address{$^1$ Departamento de F\'{\i}sica. Universidad de Oviedo.\\
Calvo Sotelo s/n. 33007 Oviedo. Spain.\\
$^2$Optics Section, Blackett Laboratory, Imperial
College,\\ London SW7 2BZ, United Kingdom}

\date{\today}
\maketitle

\begin{abstract}
The problem of the experimental determination of the amount of entanglement of a bipartite
pure state is addressed. We show that measuring a single observable does not
suffice to determine the entanglement of a given unknown pure state of two particles. 
Possible minimal local measuring strategies are discussed and a
comparison is made on the basis of their best achievable precision.
\end{abstract}

\pacs{PACS-numbers: 03.67.-a, 03.65.Bz}


\section{Introduction}
Quantum mechanical states of multiparticle systems can be entangled, a fact 
well
known since the early days of the formulation of the quantum theory \cite{gato}. 
However, the status of this property has changed substantially in recent years.
Entanglement can also be viewed as a resource and as such features in different processes
of potential practical importance, for example quantum teleportation \cite{tele}, quantum
cryptography \cite{artur} or even high precision measurements \cite{sue}. From a theoretical 
point of view, 
entanglement of bipartite pure states, and its properties
under local quantum operations, are reasonably well understood.
There is a unique measure of entanglement for these systems, provided by the
von Neumann entropy \cite{bennett}, and optimal ways
for entanglement manipulation are known \cite{varia}. However, there is a remaining practical 
question which has not been addressed so far: How can one optimally measure
the amount of entanglement of an unknown bipartite pure state?\\
At a first sight, this question may seem obvious. Reconstructing the reduced density
operator of any of the two subsystems will do the job. However, the essential
point is that we require the determination to be optimal and the
reconstruction of the reduced density matrix may provide redundant information,
given that we are asking for just a feature of the composite state, its entanglement. This is a 
single number and the first question to be answered is whether
there exists a single operator whose experimental measure may provide us with
just the amount of entanglement of the state. Note that further details of the
state itself are not of interest in the problem we are posing here \cite{ub}. We will
prove in the following that such an operator does not exist, a conclusion that
confirms what it could initially be thought as an educated guess. Knowing the impossibility
of a test using a repeated measurements of a single observable, we will discuss possible 
strategies aimed to be minimal, in the
sense of involving the smallest number of observables. In order to avoid any ambiguity
when counting the number of observables involved in a given measurement protocol, we will
define such number as the different number of {\em meters} each observer has to read out.
Among
different minimal strategies, that is, strategies involving the same number of
meters, we will call optimal the one 
providing the best accuracy when supplied with the same resources.
We will show that, in fact, measuring the reduced density matrix turns out to be 
a minimal way to proceed. Moreover, the protocol can be made optimal when
involving projections along mutually orthogonal directions.
We have organized the paper as follows. 
In section II we state formally the problem. Section III shows the impossibility
of finding a single observable whose measure may allow the experimental determination
of the amount of entanglement. Minimal local strategies are discussed in Section IV. 
When supplied with the same number of identically prepared bipartite pure states, we
discuss the performance of two classes of minimal measurements from the
point of view of the achievable precision in determining the amount of entanglement.
Section V is devoted to conclusions.
%
%
\section{An experimental scenario} 
Let us imagine the following situation. We are provided with a state preparator
which creates pairs of two-level particles (qubits) in an unknown entangled state.
These entangled pairs are distributed to two remote locations where two observers, Alice and Bob,
may perform local measurements as well as interchange classical communication. 
The internal
dynamics of the device is not specified and the only thing Alice and Bob
know is that, with high accuracy,
the state they share is pure. Therefore, the two-qubit state can be written as
\begin{equation}
\rho = \ket{\psi_{AB}} \bra{\psi_{AB}},
\end{equation}
where
\begin{equation}
\ket{\psi}_{AB} = a_0 \ket{00} + a_1 \ket{01} + a_2 \ket{10} + a_3 \ket{11}.
\end{equation}
In this expresion $(\ket{0},\ket{1})$ refer to the eigenvectors of
the operators $\sigma_z$ , the complex
coefficients
$a_i$, $\left(i=0, \dots , 3\right)$, being completely unknown.
In addition, we assume that the machine may supply a large number of identical 
pairs.
The aim is to use the resulting pairs for a quantum information task and, therefore,
the only property we are interested in is its amount of entanglement. 
Moreover, we require 
the measurement aimed to determine the amount of entanglement to be optimal in the following 
sense. First, the protocol should 
involve the smallest possible number of observables. Such tests will be called
minimal. And secondly, among minimal tests, 
we will define as optimal the class of protocols that yield the best resolution when 
supplied with the same resources, i.e., the same number of identically prepared two-level systems.\\
The problem is still rather general and, for simplicity, three further assumptions will be made:\\
\begin{enumerate}
\item The experimental situation is such that it only allows to act on one pair 
at a time. In other words, we restrict ourselves to incoherent 
measurements. Alice and Bob are not allowed to store a given number of particles and
perform a joint measurement on them \cite{barsa}.
\item No ancillary systems are available and the only allowed incoherent measurements
are projective ones. 
\item The adopted protocol is rigid, in the sense that we will not accumulate
information from a given set of initial measurements and re-adjust our 
strategy afterwards.
\end{enumerate}
In these conditions, we will show that 
no single operator measurement
allows to determine the amount of entanglement of an unknown bipartite pure state.
%
%
\section{Impossibility of a single-observable measuring strategy}
The amount of entanglement of a bipartite pure state is given by its von Neumann entropy,
\begin{equation}
E(\psi_{AB}) = - tr(\rho_A \log_2 \rho_A) = - tr(\rho_B \log_2 \rho_B)
\end{equation}
where $\rho_{A(B)}=tr_{B(A)} \rho$ is the reduced density matrix of each subsystem and 
$\rho$ is given by Eq. (1). In terms of the concurrence $C$ \cite{wooters}, defined as
\begin{eqnarray}
C^2(\psi_{AB}) &=& \vert \bra{\psi} \sigma_y \otimes \sigma_y \ket{\psi^{\ast}} \vert^2\\ \nonumber
& = & \vert a_0 a_3 - a_1 a_2 \vert^2 \\ \nonumber
& = & 4 det \, \rho_A = 4 det \, \rho_B,
\end{eqnarray}
the amount of entanglement can be expressed in a compact form as follows
\begin{eqnarray}
E(\psi_{AB}) &= - & \left(\frac{1+\sqrt{1-C^2}}{2} \right)\log_2 \left(\frac{1+\sqrt{1-C^2}}{2}\right) \\ \nonumber
& - & \left(\frac{1-\sqrt{1-C^2}}{2}\right) \log_2 \left(\frac{1-\sqrt{1-C^2}}{2}\right).
\end{eqnarray}
It should be noted that if all coefficients $a_i$ were real, the concurrence
could be obtain via the repeated measurement a single observable, $\sigma_y \otimes \sigma_y$.
We will now prove that in general, i.e. where no a priori information is
provided about the state of the bipartite system, it is not possible to evaluate
$C^2(\psi_{AB})$ by means of measuring a set of orthogonal projectors $\proy_i=
\ket{O_i} \bra{O_i}$, $\sum_{i=o}^3 \proy_i = \id$, where the $\ket{O_i}$'s form
an orthonormal basis of certain operator $\hat O$.
This measurement would allow us to compute the four probabilities 
$p_i = \vert \langle O_i \ket{\psi} \vert^2$ and therefore 
it provides three independent real numbers. It is obvious that this will not be enough to fully 
reconstruct the pure state
$\ket{\psi}$ but one may still ask whether the resulting information may be enough to 
compute a property of the state,
its amount of entanglement. In order to check this, let us first re-write the concurrence 
$C$ in a more convenient form.
For that, we will express the state $\ket{\psi_{AB}}$ in terms of the eigenbasis of the operator $\hat O$ as
\begin{equation}
\ket{\psi_{AB}} = \sum_{i=0}^3 \langle O_i \ket{\psi} \ket{O_i} \equiv \sum_{i=0}^3 m_i e^{i \phi_i} \ket{O_i},
\end{equation}
where the coefficients $m_i$'s are purely real and $\psi_i \in \left[0, 2 \pi \right)$.
Then, $C^2$ can be written as
\begin{eqnarray}
C^2(\psi_{AB})  &=& \left \vert \sum_{i=0}^3 \sum_{j=0}^3 \langle O_i \vert \psi \rangle^{\ast} \bra{O_i} 
\sigma_y \otimes \sigma_y \ket{O_j^{\ast}} \langle O_j \vert \psi \rangle^{\ast} \right \vert^2 \\ \nonumber
& = & \left \vert \sum_{i=0}^3 \sum_{j=0}^3 m_i m_j e^{-i \phi_i} e^{-i \phi_j} \langle O_i \vert \sigma_y \otimes \sigma_y
\ket{O_j^{\ast}} \right \vert^2.
\end{eqnarray}
Let us define a new matrix $K$ with elements given by
\begin{equation}
K_{ij} = \langle O_i \vert \sigma_y \otimes \sigma_y \ket{O_j^{\ast}}.
\end{equation}
In terms of this quantity, the squared concurrence can be written as 

\begin{equation}
C^2 (\psi_{AB}) = \sum_{i,j,k,l=0}^3 m_i m_j m_k m_l e^{i (\phi_k+\phi_l -\phi_i - \phi_j)} K_{ij} K_{kl}^{\ast}.
\label{concu}
\end{equation}
Looking at this expression one can already formulate the guess that it will not be possible to obtain $C^2$
from just measuring the probabilities $p_i$, given that no information about the relative phases $\psi_i$ will
be unveiled by the measurement. In what follows we will prove explicitly that $C^2$, and therefore the amount of entanglement
of the bipartite pure state, cannot be an univaluated function of the probabilities $p_i$ (equivalently, of the coefficients
$m_i$).
\subsection{An useful lemma}
Let us define two new auxiliary matrices $S$ and $\sigma$ with matrix elements given by
\begin{equation}
S_{ij} = \langle O_j^{\ast} \ket{O_i}
\end{equation}
and
\begin{equation}
\sigma_{ij} = \langle O_i  \vert \sigma_y \otimes \sigma_y \ket{O_j}.
\end{equation}
It is easy to check that the following properties hold.\\
\begin{enumerate}
\item The matrix $K$ of Eq. 8 satisfies $K=K^T$, as it follows immediately from the hermiticity of the operator
$\sigma_y \otimes \sigma_y$.
\item If the $\ket{O_i}$'s form an orthonormal basis, the corresponding conjugate vectors 
$\ket{O_i^{\ast}}$ also
form an orthonormal basis. Then, the matrix $S$ defined above is just the change of basis matrix 
between the two representations,
i.~e.,
\begin{equation}
\ket{O_i} = \sum_j S_{ij}\ket{O_j^{\ast}}, 
\end{equation}
and therefore $S^{\dagger} S =\id$ and $\vert Det(S) \vert =1$. In particular, 
this implies that $ Det (S^{\dagger}) \neq 0$.
\item $Det (\sigma)=1$.
\end{enumerate}
We now have all the ingredients for proving the following lemma.\\
{\bf Lemma:} $det(K) \neq 0$ .\\
Proof: The matrix elements of $K$ can be written in terms of those of $\sigma$ and $S$ as follows:
\begin{eqnarray*}
K_{ij} & = & \bra{O_i} \sigma_y \otimes \sigma_y \ket{O_j^{\ast}} \\ \nonumber
       & = & \sum_{l=0}^3 \bra{O_i} \sigma_y \otimes \sigma_y \ket{O_l} \langle O_l \ket{O_j^{\ast}} \\ \nonumber
       & = & \sum_{l=0}^3 \sigma_{il} S_{jl}^{\ast} = \sum_{l=0}^3 \sigma_{il} S_{lj}^{\dagger} \\ \nonumber
\end{eqnarray*}
Therefore, $K = \sigma S^{\dagger}$ and the lemma follows from properties 2 and 3.
\subsection{Impossibility of a single-observable test}
We will now prove that assuming that the measurement of a single observable allows to determine the
concurrence of the state, and therefore its amount of entanglement, yields a contradiction with the previous
lemma. Given that the state $\ket{\psi}$ is unknown, the test we are seeking must be universal, that is,
the hypothetical observable $\hat O$ has to provide the amount of entanglement of whatever input state. The
idea underlying our proof is to show that there will
always be a particular case yielding to a contradiction. Therefore, if no a priori information is provided,
the minimal test will necessarily require measuring more than one observable.\\
Consider the particular case where $m_0=m_1=1/\sqrt{2}$ and $m_2=m_3=0$. In this case, Eq. \ref{concu}
takes the form:
\begin{eqnarray*}
C^2 &=& \frac{1}{4} ( \vert K_{00} \vert^2 + \vert K_{11} \vert^2 + 4 \vert K_{01} \vert^2 \\
&+& 2 e^{i \phi} ( K_{00}K_{01}^{\ast} + K_{01}K_{11}^{\ast}) + 
2 e^{-i \phi}( K_{01}K_{00}^{\ast} + K_{11}K_{01}^{\ast})\\
&+& e^{2 i \phi} K_{00}K_{11}^{\ast} + e^{- 2 i \phi} K_{11}K_{00}^{\ast} )\\
\end{eqnarray*}
where we have called $\phi=\phi_1-\phi_0$ and had make use of property 1. If we assume that $C^2$ is
only a function of the real numbers $m_i$, i. e. independent of the relative phase $\phi$, the fact that
the functions $(1, e^{i \phi}, e^{-i \phi}, e^{2 i \phi}, e^{-2 i \phi})$ are linearly
independent yields the set of equalities
\begin{equation}
C^2 = \frac{1}{4} ( \vert K_{00} \vert^2 + \vert K_{11} \vert^2 + 4 \vert K_{01} \vert^2) 
\label{ka}
\end{equation}
\begin{equation}
K_{00}K_{01}^{\ast} + K_{01}K_{11}^{\ast}=0
\label{ka1}
\end{equation}
\begin{equation}
K_{00}K_{11}^{\ast}=0
\label{ka2}
\end{equation}
Eq. \ref{ka2} implies that either $K_{00}=0$ and/or $K_{11}=0$. Taking Eq. \ref{ka1} into account,
this corresponds to the cases where $K_{01} \neq 0$ or $K_{01} = 0$. In other words, we obtain that
two among the three complex numbers $(K_{00},K_{01},K_{11})$ must be zero.\\
If we repeat this argument for all the cases where any two of the coefficients $m_i$ are
equal to $1/\sqrt{2}$ and the remaining two equal to zero, we end up with the requirement that
in all the following sets of three complex numbers 
$$
\begin{array}{ccc}
(K_{00},K_{01},K_{11}) & (K_{00},K_{02},K_{22}) & (K_{00},K_{03},K_{33})\\
(K_{11},K_{12},K_{22}) & (K_{11},K_{13},K_{33}) & \\
(K_{22},K_{23},K_{33}) &                        &
\end{array}
$$
there must be at least two of them equal to zero in any set. This fact imposes a certain symmetry
for the allowed $K$-matrices. Explicitly, $K$ can only be one of the following
\begin{equation}
	K_1 = \left(
	\begin{array}{cccc}
	 K_{00} & 0 & 0      & 0     \\
	 0      & 0 & K_{12} & K_{13}\\
	 0      & K_{12} & 0 & K_{23}\\
       0      & K_{13} & K_{23} & 0
	\end{array}\right) 
\end{equation}
or
\begin{equation}
	K_2 = \left(
	\begin{array}{cccc}
	 0 & K_{01} & K_{02} & 0     \\
	 K_{01} & 0 & K_{12} & 0 \\
	 K_{02}   & K_{12} & 0 & 0\\
         0   & 0 & 0 & K_{33}
	\end{array}\right), 
\end{equation}
and analogous forms obtained when interchanging the roles of the indexes, or of the form
\begin{equation}
	K_3 = \left(
	\begin{array}{cccc}
	 0 & K_{01} & K_{02} & K_{03}     \\
	 K_{01} & 0 & K_{12} & K_{13} \\
	 K_{02}   & K_{12} & 0 & K_{23}\\
         K_{03}   & K_{13} & K_{23} & 0
	\end{array}\right), 
\end{equation}
It should be noted that many other cases could be obtained if any of the matrix elements written as non-zero were zero,
however these additional cases are not of interest here, as will become clear afterwards.\\
Our proof ends with showing that in any of the allowed forms for $K$, some of the possibly nonzero coefficients in K turn out to be zero. Therefore, all the allowed forms for $K$, i.e. all forms compatible with the requirement of being the concurrence a univaluated
function of the real numbers $p_i$, will have determinant equal zero, which contradicts the lemma stated before.\\
This can be easily shown  for matrices of the form $K_1$ or $K_2$ just following an argument parallel to one used above and 
choosing three of the coefficients $m_i$ equal to $1/\sqrt{3}$ and the remaining one equal to zero. Let us analyze here the
case of $K$-matrices of the form $K_3$. If we set the $m_i$'s coefficients to the values $m_0=0$ and
$m_i=1/\sqrt{3}$ for $(i=1,...,3)$, the squared concurrence given by Eq. \ref{concu} reads
\begin{eqnarray*}
C^2 &=& \frac{4}{3 \sqrt{3}} ( \vert K_{12} \vert^2 + \vert K_{13} \vert^2 + \vert K_{23} \vert^2 \\
& + & e^{i \alpha} K_{12} K_{13}^{\ast} + e^{-i \alpha} K_{13} K_{12}^{\ast}\\
& + & e^{i \beta}  K_{12} K_{23}^{\ast} + e^{-i \beta}  K_{23} K_{12}^{\ast}\\
& + & e^{i \gamma} K_{13} K_{23}^{\ast} + e^{-i \gamma} K_{23} K_{13}^{\ast} )
\end{eqnarray*}
where we have introduced the relative phases $\alpha=\phi_3-\phi_2$, $\beta=\phi_3-\phi_1$ and $\gamma=\phi_2-\phi_1$.
Using again the argument invoked in proving the allowed forms for the matrix $K$, we obtain that the following equalities must hold
$$
\begin{array}{c}
K_{12} K_{13}^{\ast}=0 \\ 
K_{12} K_{23}^{\ast}=0 \\
K_{13} K_{23}^{\ast}=0.
\end{array}
$$
Therefore, within the three numbers $(K_{12},K_{13},K_{23})$, two must be zero. If we repeat the argument making each time one of the
$m_i$'s zero and the other three equal to $\sqrt{3}$, we end up with allowed forms for $K_3$ 
with either four or two matrix elements different from zero.
For instance, an allowed form of $K$ is given by
\begin{equation}
	K_3^1 = \left(
	\begin{array}{cccc}
	 0 & K_{01} & 0 & 0     \\
	 K_{01} & 0 & 0 & 0 \\
	 0 & 0 & 0 & K_{23}\\
         0 & 0 & K_{23} & 0
	\end{array}\right).
\end{equation}
But choosing now all the coefficients $m_i$ equal to $1/2$ suffices to get the constraint $K_{01}K_{23}^{\ast}=0$,
which yields a zero determinant for $K$. The same reasoning
applies to the other five possible cases. Therefore, assuming that $C^2$ is only a function of the probabilities $p_i$ yields to
the condition $det(K)=0$, but we have proved in the previous section that the orthonormality of the vectors $O_i$ demands the determinant
of $K$ to be non zero. As a result, it is not possible to find a single operator $\hat O$ whose measurement allows to determine
the amount of entanglement of the pure state $\ket{\psi_{AB}}$.
%
%

\section{Minimal tests}
The previous analysis shows that a measuring strategy employing a single-observable does not allow Alice and Bob to know the 
amount of entanglement of the state they are sharing. 
Knowing this fact, the natural question to ask is to determine the minimal measuring
strategy than may allow them to evaluate $E(\psi_{AB})$.
It is clear that if they measure two different observables $\hat O_{AB}$ of the form analyzed before, 
they will fully reconstruct the original pure state and can, therefore, compute its amount of 
entanglement. It is a remarkable fact that acting on the whole Hilbert
space of the two particles, 
we cannot isolate the information related to the amount of entanglement alone by means of measuring a 
single observable. If no information is known about the state, determining its amount of entanglement 
leads to a full reconstruction of the state. 
However, such a non-local implementation may not be
the easiest to implement experimentally and local strategies are preferred. We will discuss in this 
section possible ways to proceed if
one is constrained to act locally and we will determine the expected precision of the protocols.

\subsection{Local actions without exchange of classical communication}
When Alice and Bob are constrained to act locally and no classical communication can be exchanged among 
them, the minimal measuring strategy corresponds to the local reconstruction of the reduced density 
operators. For instance, Alice may reconstruct
the operator $\rho_A=tr_B \rho_{AB}$ and send at the end of the protocol a final message to Bob whose 
length will depend on the required resolution.\\
To achieve this, Alice needs to perform three projective measurements along linearly independent directions. We will 
show in the following that measuring along three orthogonal directions is in fact optimal, in the sense that 
choosing this configuration yields the smallest associated uncertainty in the
experimental determination of the determinant of the reduced density operator (See Eq. (4)).\\
Let us write the reduced density matrix in the general form
\begin{equation}
	\rho_A = \frac{1}{2} (\id + \sigma. S) = \frac{1}{2} \left(
	\begin{array}{cc}
	 1+S_z & S_x - i S_y \\
         S_x + i S_y  & 1- S_z 
	\end{array}\right).
\end{equation}
in terms of the corresponding Bloch's vector.
With the above parametrization we have a one to one 
correspondence between directions in three dimensional space and directions within the Bloch sphere. Note that the determinant of $\rho_A$ only depends on the modulus of Bloch's vector. In other words, it 
is rotationally invariant. Suppose now we are planning to measure the amount of entanglement projecting 
the reduced density matrix of a given state along three linearly independent directions. 
The uncertainty associated with this measurement will depend on:
\begin{enumerate}
\item the modulus of the corresponding Bloch's vector, as the amount of entanglement does.
\item the relative position of Bloch's vector with respect to the three projective directions.
\end{enumerate}
Because of conditions 1 and 2, 
assuming the initial distribution 
of states to be isotropic, the average uncertainty after measuring 
sufficiently many states will only depend on the relative position of the three directions we project along.
In particular, this implies we can choose a given direction to be the $z$-axis. We will call 
$\hat n$ and $\hat m$ the other two directions so that the angles they form with the $z$-axis are $\theta_{n}$ 
and $\theta_{m}$. Then we can write the average uncertainty as
\begin{center}
 $\delta_{av}=f(\theta_{n},\theta_{m},\phi_{nm}).$
\end{center}
Here, $\phi_{nm}$ is the relative azimut angle $\phi_m-\phi_n$. Moreover, 
because of condition 2, the following equalities must hold
$$ f(\theta_{n},\theta_{m},\phi_{nm})=f(\pi-\theta_{n},\theta_{m},\pi-\phi_{nm}),$$
$$ f(\theta_{n},\theta_{m},\phi_{nm})=f(\theta_{n},\pi-\theta_{m},\phi_{nm}-\pi),$$
$$ f(\theta_{n},\theta_{m},\phi_{nm})=f(\pi-\theta_{n},\pi-\theta_{m},\phi_{nm}).$$
This is equivalent to say that we could redefine the three positive axis (simultaneously or not) 
without changing the average uncertainty. Finally, from the previous set of equations one obtains
that the function $f$ has to have an extremum at $$ \theta_{m}=\pi/2 , \theta_{n}=\pi/2 , \phi_{nm}=\pi/2. $$ 
We have numerically calculated the average uncertainty $\delta _{av}$ over 10.000 states of the 
composite system uniformly distributed over the complex four dimensional joint Hilbert space and calculated 
this uncertainty for a series of different measurements with  $\theta_{m} = \pi/2, \theta_{n}=\pi/2$ 
but  $\phi_{nm}$ ranging from 0 to $\pi$.
Results are shown in Figure 1, where we show the mean uncertainty in determining the amount of entanglement $\delta_{av}$, defined as
\begin{equation}
\delta_{av}  = \left< \sqrt{
\left\vert \frac{\partial det(\rho_a)}{\partial P_0} \right \vert^2 \delta P_0 + 
\left \vert \frac{\partial det(\rho_a)}{\partial P_m} \right \vert^2 \delta P_m + 
\left \vert \frac{\partial det(\rho_a)}{\partial P_n} \right \vert^2 \delta P_n } \right>.
\end{equation}
as a function of the relative difference $\phi_{nm}$ for a fixed value of $\theta_m=\theta_n$. 
In the definition above, $P_0$, $P_m$ and $P_n$ are the probabilities to obtain a {\em spin up} 
when measuring along directions
$\hat z$, $\hat m$ and $\hat n$ respectively and
where each $\delta P$ denotes the squared variance given by
$\delta P = \frac{P (1-P)}{N}$. The bracket means the average over the isotropic distribution of states.
It is clear that $\delta _{av}$ reaches in fact a minimum when 
$\phi_{nm}=\pi/2$. Similar figures can be plotted, all of them supporting that 
the minimum uncertainty is indeed achieved when the three directions of projection are chosen to be mutually 
orthogonal.
\begin{figure}[htb]
\begin{center}
\epsfxsize11.0cm
\centerline{\epsfbox{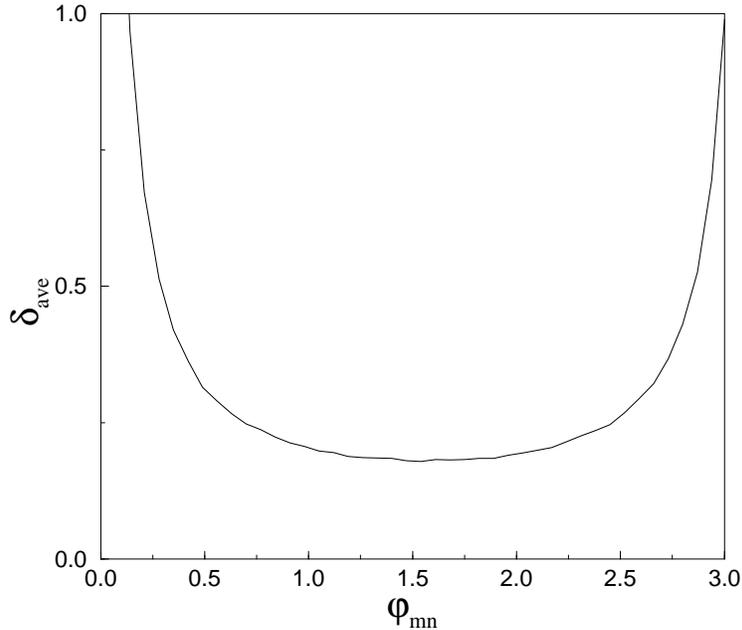} }
\label{fig1}
\end{center}
\caption{\small Average uncertainty as a function of the relative azimut angle $\phi_{mn}$ for a fixed
value of $\theta_m=\theta_m$. The average uncertainty in determining the amount of entanglement is minimal when the three 
linearly independent
directions of projection are chosen to be orthogonal.}
\label{f:ent}
\end{figure}

\subsection{Local actions with exchange of classical communication}
Let us now assume that Alice and Bob agree to cooperate. 
Then, the amount of entanglement can be evaluated from the measurement of two Pauli operators in each side. 
If they agree to measure different operators in each round, they again fully reconstruct the state. 
However, 
if one of them always 
measures the same Pauli operator, for instance they choose to compute the observables
$\sigma_z \otimes \id$ and $\id \otimes \sigma_z$ and, in a subsequent round, the observables
$\sigma_x \otimes \id$ and $\id \otimes \sigma_z$, for which they should read out three {\em meters},
they can obtain the amount of entanglement
but they will neither get full information about the state itself, nor about the whole reduced density
matrix. 
Indeed, if we denote by $P_i$ (i=0,1,2,3) the four probabilities associated to the outcomes
$(++$,$+-$,$-+$ and $--)$ when measuring 
$\sigma_z \otimes \id$ and $\id \otimes \sigma_z$ 
and 
by $P_{++}$,$P_{+-}$,$P_{-+}$ and $P_{--}$ the corresponding probabilities when measuring
$\sigma_x \otimes \id$ and $\id \otimes \sigma_z$, it can be easily shown that the probabilities $P_{ij}$ ($i,j=+,-$),
can be written in terms of the probabilities $P_i$ as follows:
\begin{eqnarray}
P_{++} &=& \frac{1}{2} \left( P_0 + P_1 + 2 \sqrt{P_0 P_1} \cos \phi_{01} \right) \\ \nonumber
P_{+-} &=& \frac{1}{2} \left( P_0 + P_1 - 2 \sqrt{P_0 P_1} \cos \phi_{01} \right) \\ \nonumber
P_{-+} &=& \frac{1}{2} \left( 1 - P_0 - P_1 + 2 \sqrt{P_3 P_2} \cos \phi_{23} \right) \\ \nonumber
P_{--} &=& \frac{1}{2} \left( 1 - P_0 - P_1 - 2 \sqrt{P_3 P_2} \cos \phi_{23} \right),
\end{eqnarray} 
where we have rewritten the amplitudes of the initial state as $a_i = m_i e^{i \phi_i}$, (i=0,1,2,3),
and called $\phi_{ij}=\phi_i-\phi_j$. From the previous set of equations, we see that the
functions $\cos(\phi_0-\phi_1)$ and $\cos(\phi_2-\phi_3)$ can be expressed in terms of measurable
quantities in the form
\begin{eqnarray}
\cos(\phi_0-\phi_1)&=& \frac{2 P_{++} - P_0 - P_1}{2 \sqrt{P_0 P_1}}\\ \nonumber
\cos(\phi_2-\phi_3)&=& \frac{2 P_{-+} + P_0 + P_1}{2 \sqrt{P_3 P_2}}.
\end{eqnarray}
Noting that 
\begin{equation}
C^2  = 4 \,\left(P_1 P_2 + P_0 P_3 - 2 \sqrt{P_0 P_1 P_2 P_3} \cos(\phi_0-\phi_1+\phi_3-\phi_2)\right).
\end{equation} 
we see that this measuring strategy suffices to determine the squared concurrence and 
therefore the amount of entanglement of the pure state but it does not allow the
full reconstruction of the initial state.
As we will show in the next section, this protocol is not optimal, in the sense of 
providing the best accuracy when measuring locally a minimal set of observables.\\
\subsection{Which strategy yields the best resolution?}
The measuring strategies described above are both minimal, in the sense of involving the smallest
number of meters to be read out.
However, it is not obvious whether the precision achieved following these two strategies
will be the same.  In fact, we will show in the following that it is not. When provided with the same 
resources, that is, using the same number of identically prepared entangled pairs, we can get the amount of 
entanglement with higher precision by means of the local reconstruction of
the reduced density operator. If we denote by $N$ the number of entangled pairs, $N$ being large in the
statistical sense, a numerical simulation with $10^6$ states from an isotropic initial distribution
yields the following results.\\
\begin{enumerate}
\item The measurement procedure by means of the local reconstruction of the reduced density operator 
has an associated uncertainty which scales with $N$ as
$$ \delta_{loc} = \frac{0.3}{\sqrt{N}}. $$
\item The associated uncertainty with a local measurement of the form described in Sec. IV.B is 
substantially much larger \cite{sime}.
More precisely
$$ \delta_{loc+cc} = \frac{2.3}{\sqrt{N}}. $$
\end{enumerate}
Note that, once $N$ is given, the resulting number of measurements in each measuring protocol is different.
While in the first case each single probability will scale as $P \approx 1/\sqrt{N/3}$, the larger number
of measurements yields each probability in the second procedure to scale as $P \approx 1/\sqrt{N/2}$.\\
From these results one may be lead to the conclusion that the best resolution will always be achieved by means
of reconstructing the reduced density operator. However, this may not be true. Imagine that, in the context
of the second protocol, Bob measures a different Pauli operator. If the direction of projection is orthogonal
to the $z$-axis, this procedure will also allow to reconstruct the initial state. Will now the associated
uncertainty be reduced with respect to the case analyzed above? In the light of the results obtained when
measuring the reduced density operator, this should be the case. It should be noted, however, that the
number of observables required in this measurement protocol is no longer minimal, as it would require
the observer Bob to read out an additional meter. 
\section{Conclusions}
We have analyzed the problem of determining experimentally the amount of entanglement of bipartite
pure states when one has a large supply of identically prepared systems on which one is restricted to
act by means of projective measurements. We showed that, provided that the entangled state is totally
unknown, no measuring strategy involving a single operator exists. Therefore, acting on the Hilbert space
of the composite system does not allow to single out the amount of entanglement without allowing to
determine the state completely. When local actions are considered, the minimal protocol for determining
the amount of entanglement
involves measuring three different observables. We have analyzed here two classes of minimal tests. 
In the first one, no exchange of classical information is required and entails the local reconstruction of 
the reduced density operator. The procedure is optimal,
in the sense of having the smallest associated uncertainty, when measuring along three mutually orthogonal
directions. 
The second class of protocols requires the use of classical information. 
Here we have
analyzed a possible strategy and showed that it suffices to determine the amount of entanglement of the
pure state but not its full reconstruction. The associated resolution turns out to be worse than the
one corresponding to the measurement of the reduced density operator. The analyzed protocol is not
necessarily the most precise among the whole class of measuring strategies by means of local actions
with the exchange of classical information. However, an increase in the resolution would be done at the prize of
increasing the number of meters to be read out and the protocol would no longer be minimal. 
Establishing how the best accuracy achievable within a protocol 
of this type, and which allows the full reconstruction of the state,
compares with the one associated to the reconstruction of the the reduced density operator is an interesting open problem.

{\bf Acknowledgements:} The authors thank M.B.~Plenio, P. Hayden, D. Jonathan, G. Vidal, A.K.~Ekert, 
D.P~DiVincenzo and J.I.~Cirac for discussions on 
the subject of this paper. J.M.G.S. also acknowledges M. Ferrero for continuous encouragement and 
useful discussions. This 
work has been supported by The Leverhulme Trust, the European 
Science Foundation, The Engineering and Physical Sciences Research Council (EPSRC) and DGICYT Project No. 
PB-95-0594 (Spain) and has benefited from the participation in the ESF-QIT workshop in Cambridge.


\begin{thebibliography}{99}
\bibitem{sime} The same resolution would be obtained if Alice proyects along directions $\hat x$ and $\hat y$
or along $\hat z$ and $\hat y$.
\end{thebibliography}
\end{document}